\documentclass{jpp}
\usepackage[utf8]{inputenc}
\usepackage[english]{babel}
\usepackage[T1]{fontenc}
\usepackage{lmodern}
\usepackage{amsmath}
\usepackage{graphicx}
\usepackage{subcaption}
\usepackage[usenames,dvipsnames]{color}
\usepackage{amssymb}
\usepackage{verbatim}
\usepackage{algorithm}
\usepackage{algpascal}
\usepackage{hyperref}
\usepackage{listings}
\usepackage{units}
\usepackage{fancyhdr}

\makeatletter

\newsavebox\myboxA
\newsavebox\myboxB
\newlength\mylenA

\newcommand*\obar[2][0.75]{
    \sbox{\myboxA}{$\m@th#2$}%
    \setbox\myboxB\null
    \ht\myboxB=\ht\myboxA%
    \dp\myboxB=\dp\myboxA%
    \wd\myboxB=#1\wd\myboxA
    \sbox\myboxB{$\m@th\overline{\copy\myboxB}$}
    \setlength\mylenA{\the\wd\myboxA}
    \addtolength\mylenA{-\the\wd\myboxB}%
    \ifdim\wd\myboxB<\wd\myboxA%
       \rlap{\hskip 0.5\mylenA\usebox\myboxB}{\usebox\myboxA}%
    \else
        \hskip -0.5\mylenA\rlap{\usebox\myboxA}{\hskip 0.5\mylenA\usebox\myboxB}%
    \fi}

\makeatother

\renewcommand{\vec}[1]{\boldsymbol{#1}}

\renewcommand{\d}{\ensuremath{\mathrm{d}}}

\newcommand{\lang}{\left\langle}
  \newcommand{\rang}{\right\rangle}
\newcommand{\stella}{\textsc{Stella}}
\newcommand{\GX}{\textsc{GX}}

\newcommand{\nouncite}[1]{\citet{#1}}

\newcommand{\VMEC}{\textsc{Vmec}}
\newcommand{\DESC}{\textsc{Desc}}

\newcommand{\simsopt}{\textsc{Simsopt}}
\newcommand{\scipy}{\textsc{scipy}}

\usepackage{stackengine}
\stackMath

\addto\extrasenglish{ 	 	 	 }

\shorttitle{Quasi-axisymmetric stellarators with varied rotational transform}
\shortauthor{S.~Buller, M.~Landreman, J.~Kappel, R.~Gaur}

\title{A family of quasi-axisymmetric stellarators with varied rotational transform}

\author{S.~Buller\aff{1}
  \corresp{\email{sbuller@umd.edu}},
  M.~Landreman\aff{1},
  J.~Kappel\aff{1},
R.~Gaur\aff{1}}

\affiliation{\aff{1}Institute for Research in Electronics and Applied Physics, University of Maryland, Maryland, MD 20742, USA}

\begin{document}

\maketitle

\begin{abstract}
  We apply a continuation method to recently optimized stellarator equilibria with excellent quasi-axisymmetry (QA) to generate new equilibria with a wide range of rotational transform profiles.
  Using these equilibria, we investigate how the rotational transform affects fast-particle confinement, the maximum coil-plasma distance, the maximum growth rate in linear gyrokinetic ion-temperature gradient (ITG) simulations, and the ion heat flux in corresponding nonlinear simulations. We find values of two-term quasisymmetry error comparable to or lower than the similar Landreman-Paul (Phys. Rev. Lett. \textbf{128}, 035001) configuration for values of the mean rotational transform $\bar{\iota}$ between $0.12$ and $0.75$.
  The fast-particle confinement improves with $\bar{\iota}$ until $\bar{\iota} = 0.73$, at which point the degradation in quasisymmetry outweighs the benefits of further increasing $\bar{\iota}$. The required coil-plasma distance only varies by about $\pm 10\%$ for the configurations under consideration, and is between $2.8\,\mathrm{m}$ to $3.3\,\mathrm{m}$ when the configuration is scaled up to reactor size. The maximum growth rate from linear gyrokinetic simulations increases with $\bar{\iota}$, but also shifts towards higher $k_y$ values. The maximum linear growth rate is sensitive to the choice of flux tube at rational $\iota$, but this can be compensated for by taking the maximum over several flux tubes. The corresponding ion heat fluxes from nonlinear simulations display a non-monotonic relation to $\iota$.
  Sufficiently large positive shear is destabilizing. This is reflected both in linear growth rates and nonlinear heat fluxes.
\end{abstract}

\section{Introduction\label{sec:intro}}
Stellarators are toroidal magnetic confinement devices that rely on non-axisymmetric magnetic fields to confine a plasma. Non-axisymmetry makes it possible to produce the confining magnetic field using currents external to the plasma. However, without an explicit symmetry direction, particles are not guaranteed to be confined by conservation laws, and the magnetic field must be optimized to confine particles.

One possible optimization strategy is to optimize for a property known as quasisymmetry \citep{boozer1983}, in which the magnitude of the magnetic field has a symmetry in a special set of coordinates.
Over the last few years, tremendous progress has been made in optimizing stellarator geometries to achieve excellent quasisymmetry \citep{landremanPaul2022}.
However, the results so far are mostly limited to configurations scattered in parameter space, which makes it hard to systematically study the effects of varying equilibrium parameters, such as the rotational transform $\iota$ and global magnetic shear $\hat{s}$. 

In this paper, we attempt to alleviate this issue by applying continuation methods to continuously deform the rotational transform profile of the ``Precise QA'' vacuum configuration described in \nouncite{landremanPaul2022}.
This procedure lets us generate a family of stellarator geometries that are similar to each other in a sense that will be explained in the next section, but where the rotational transform profile varies.

Since the mean rotational transform of the original Precise QA configuration was arbitrarily imposed, it is not a priori obvious that there are any inherent trade-offs between quasisymmetry and different values of mean $\iota$. Increasing the rotational transform is expected to improve the fast particle confinement, since the width of banana orbits scales $\propto 1 / \iota$, \citep{paul2022}, and could thus be beneficial for a reactor.


Optimization by continuation is a technique whereby an optimization problem is solved by continuously deforming a simpler optimization problem.
If each deformation is a small perturbation to the previous problem, the solution of the previous problem can be used as a good initial guess to the perturbed problem.
By repeatedly applying small perturbations and optimizing at each step, we eventually arrive at a solution to the novel optimization problem.

A variety of continuation-type methods have been used for stellarator equilibrium and optimization calculations previously.
The code \DESC{} allows for MHD equilibria to be computed by continuation in the plasma boundary shape or plasma profiles \citep{conlin2023}.
One common optimization approach -- optimizing first in a small parameter space of Fourier modes describing the plasma shape, then expanding the number of Fourier modes in the parameter space and re-optimizing -- can be viewed as a type of continuation method.
This approach was discussed for instance in \citet{landremanPaul2022}.
Continuation is also a standard method for exploring the Pareto front in multi-objective optimization, and has recently seen use in this context to explore the trade-offs between quasisymmetry and aspect ratio \citep{padidar2023}.

Larger systematic explorations of the space of quasisymmetric stellarators have previously been published by \citet{landreman_rainbow2022}, which relied on the near-axis expansion to directly construct optimized stellarator equilibria  \citep{landreman_sengupta_plunk_2019, plunk_landreman_helander_2019}. In contrast, our geometries are solutions to the full equilibrium equation, with a realistic number of Fourier modes describing the boundary.
\citet{giuliani2023} recently made public a large database of QA stellarators with varying $\iota$, which could be used to perform studies similar to this one.



The rest of the paper is organized as follows.
In the next section, we present the objective function used in the optimization, and introduce the continuation method. In \autoref{sec:results}, we apply this method to generate ``Precise QA''-like configurations with varying mean rotational transform or shear. We analyze these configurations with respect to fast-particle confinement, gyrokinetic stability, and required coil-plasma distance. We find that fast-particle confinement improves with higher $\iota$, up to a point, while the maximum growth rate in our gyrokinetic simulations increases and shifts towards higher $k_y$.
Finally, we discuss the results in \autoref{sec:discussion}. All the equilibria presented here can be found as supplemental material on Zenodo \url{https://zenodo.org/doi/10.5281/zenodo.10521393}, see \citet{buller_2024}.

\section{Theory\label{sec:theory}}
A magnetic field $\vec{B}$ is quasisymmetric with helicity $(M,N)$ if its magnitude $B = |\vec{B}|$ can be written as
\begin{equation}
B = B(\psi, M\vartheta - N\zeta),
\end{equation}
where $\psi$ is a flux-surface label, which we take to be the toroidal flux divided by $2\pi$; and $\vartheta$ and $\zeta$ are magnetic angles, which we take to be the Boozer angles.
Compared to the generic case where $B = B(\psi,\vartheta,\zeta)$ depends on all angles independently, the Lagrangian of a charged particle traveling through a quasisymmetric field has a symmetry direction when written in Boozer coordinates, which ensures that the charged particle is confined by the magnetic field \citep{boozer1983,nuhrenberg1988,rodriguez2020necessary}. The symmetry direction depends on $M$ and $N$: fields with $N=0$ are known as quasi-axisymmetric (QA); fields with nonzero $N$ and $M$ are called quasihelically (QH) symmetric. 

As the Boozer angles $\vartheta$ and $\zeta$ themselves depend on $\vec{B}$, it is not known how to directly compute fields with quasisymmetry, except for the case where the fields possess exact axisymmetry ($\vec{B} = \vec{B}(\psi,\vartheta)$). Using different asymptotic methods for finding quasisymmetry typically yields overdetermined systems of equations beyond a certain order \citep{garrenBoozer2, plunk_helander_2018}, which suggests that exact quasisymmetry apart from true axisymmetry may be impossible to achieve.
The approach taken historically is to instead optimize $\vec{B}$ to achieve approximate quasisymmetry, which can yield configurations that are good enough in practice.
This is also the approach taken in this paper.


As in \nouncite{landremanPaul2022}, we take our objective function to be
\begin{equation}
f = f_{\text{QS}} + (A - A_*)^2 +   (\bar{\iota} - \bar{\iota}_*)^2,\label{eq:objective}
\end{equation}
where $A$ is the aspect ratio, $\bar{\iota}$ the mean (over $\psi$) of the rotational transform, and $f_{\text{QS}}$ is the so-called \emph{two-term} formulation of quasisymmetry \citep{helanderSimakov2008}. Specifically
\begin{equation}
f_{\text{QS}} = \sum_{s_j} \lang \left(\frac{1}{B^3} [N - \iota M] \vec{B} \times \nabla B \cdot \nabla \psi - [MG + NI] \vec{B} \cdot \nabla B\right)^2\rang, \label{eq:2term}
\end{equation}
where the sum is over flux surfaces, $\langle \cdot \rangle$ denote the flux-surface average, $2\pi G/\mu_0$ and $2\pi I/\mu_0$ are the poloidal current outside and toroidal current inside the flux surface, respectively. The $f_{\text{QS}}$ objective is zero when $\vec{B}$ is quasisymmetric with helicity $(M,N)$, so different kinds of quasisymmetry can be targeted by specifying $M$ and $N$. The second and third term on the right-hand side of \eqref{eq:objective} target a specific aspect ratio $A_*$ and mean rotational transform $\bar{\iota}_*$, respectively.

The above objective is optimized with respect to the boundary shape of the plasma, which is described in terms of Fourier modes. The boundary can be described by
\begin{equation}
        \begin{aligned}
        R(\theta, \phi) =& \sum_{m=0}^{M\text{pol}}
          \sum_{n=-N\text{tor}}^{N\text{tor}} r_{c,m,n} \cos{(m \theta - n_{\text{fp}} n \phi)}, \\
          Z(\theta, \phi) =& \sum_{m=0}^{M\text{pol}}
          \sum_{n=-N\text{tor}}^{N\text{tor}} z_{s,m,n} \sin{(m \theta - n_{\text{fp}} n \phi)},
        \end{aligned}
\end{equation}
where $R$, $Z$, $\phi$ are the cylindrical coordinates of our surface and $M_{\text{pol}}$ and $N_{\text{tor}}$ are the maximum poloidal and toroidal mode numbers in our representation. Throughout this work, stellarator symmetry is assumed, which eliminates the sine and cosine components in $R$ and $Z$, respectively. The poloidal angle $\theta$ is arbitrary.

For each boundary, a magnetic equilibrium is calculated using the \VMEC{} code \citep{vmec1983} and the objective is evaluated for this calculated equilibrium. The optimization is formulated using the \simsopt{} framework \citep{simsopt}. We use the Trust Region Reflective method in \scipy{} to find a local minimum given an initial condition \citep{2020SciPy-NMeth}.

\subsection{Computational procedure\label{sec:cp}}
We generate our equilibria by solving a series of local optimization problems with different $\bar{\iota}_*$, using a \emph{continuation method}.

Let $F[\bar{\iota}_*]$ denote the objective function \eqref{eq:objective} with a specific value of $\bar{\iota}_*$.
Let $F^{(0)} = F[\bar{\iota}_*^{(0)}]$ and let $\vec{x}^{(0)}$ be an optimum to this objective. We define our series of optimization problems through
\begin{align}
  \bar{\iota}_*^{(n+1)} =& \bar{\iota}_*^{(n)} + \Delta \bar{\iota}_* \\
  F^{(n+1)} =& F[\bar{\iota}_*^{(n+1)}];
\end{align}
and take $\vec{x}^{(n)}$ to denote the optimum of $F^{(n)}$.

Since we are doing local optimizations, the obtained optimum for a given $F^{(n)}$ will depend on the initial guess.
For each new optimization problem $F^{(n+1)}$, we use the optimum from the previous optimization problem $F^{(n)}$ as our initial guess. If $\Delta \bar{\iota}_*$ is small, we expect $F^{(n+1)}$ to be roughly equal to $F^{(n)}$ with a small perturbation, and $\vec{x}^{(n)}$ can thus be expected to be close to the new optimum $\vec{x}^{(n+1)}$. If this expectation holds, our procedure will generate a series of equilibria with different $\bar{\iota}_*$ that are otherwise similar, in the sense of being close to each other in the space of Fourier boundary modes which we optimize over.

\section{Results\label{sec:results}}
In this section, we present measures of quasisymmetry, fast-particle confinement and ion-temperature-gradient (ITG) stability for the different geometries obtained using the procedure described in \autoref{sec:cp}. Quasisymmetry is part of the objective function \eqref{eq:objective}, and can thus be considered a measure of how well the optimizer performs for the different $\bar{\iota}_*$. The other figures of merit were not explicitly optimized for.

\subsection{Exploring the optimization landscape around an optimum\label{sec:explore}}
\begin{figure}
  \includegraphics[width=\textwidth]{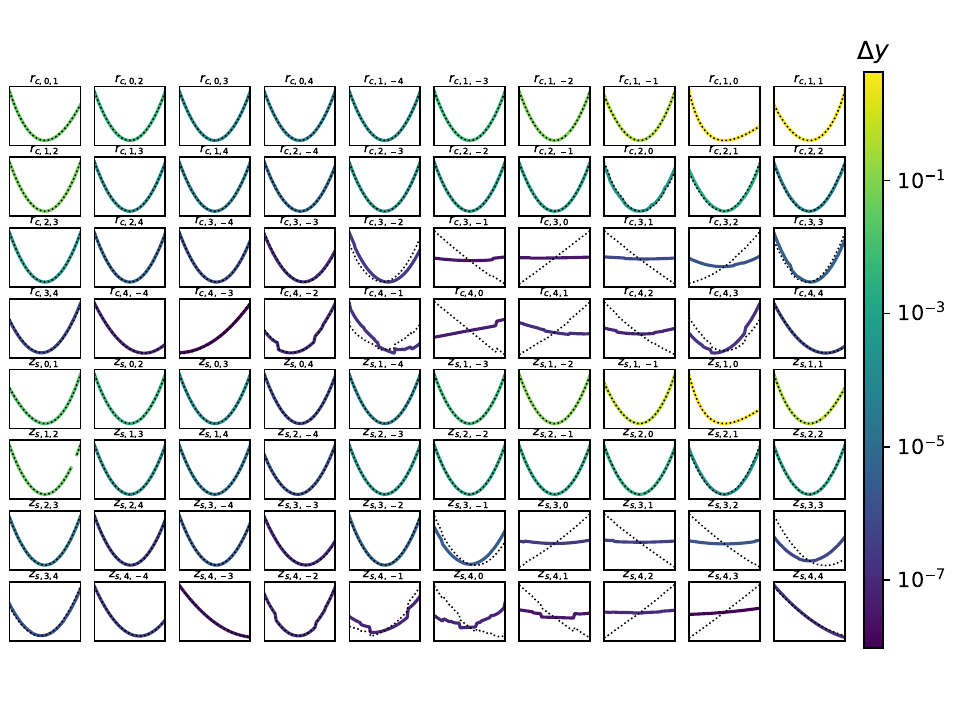}
  \caption{\label{fig:landscape1} Solid lines: Variation in the objective \eqref{eq:objective} for $\bar{\iota}_*=0.42$ for a scan in boundary modes around the corresponding optimum. Each subplot corresponds to a scan in one boundary mode, as indicated above the subplot. Each boundary mode is varied by up to $\pm 30\%$ and the colors correspond to the range of values taken by the objective during this scan. Dashed lines: objective evaluated for the same boundaries, but with $\bar{\iota}_*=0.43$.}
\end{figure}

Before presenting detailed results of our optimized configurations, it is instructive to look at the optimization landscape around an optimum and how it changes when going to the next $\bar{\iota}_*$.
In \autoref{fig:landscape1}, we plot the objective function when varying each of the Fourier boundary modes about our optimum for $\bar{\iota}_* = 0.42$. Each subfigure corresponds to a different boundary mode, indicated above the plot. The $y$-axes and $x$-axes correspond to the value of the objective and the boundary mode, respectively. Color is used to indicate the range of the $y$-axis. The range of the $x$-axis covers a  $30\%$ relative change in the corresponding boundary parameter.

From reading the colorbar, we see that the relative sensitivity of the objective to various boundary modes varies by almost 8 orders of magnitude. This is largely due to the differences in the absolute values of the original modes. Since several boundary modes have very shallow or even non-existent local wells, our optimum is effectively in a flat valley and is thus not unique. Small changes in the numerical optimization method or initial condition will result in optima where the non-sensitive boundary modes assume different values, within the narrow range plotted. 

This ``non-uniqueness'' suggests that the slight perturbations of the continuation method may cause unpredictable jumps in the insensitive boundary modes. 
To illustrate the immediate effect of increasing $\bar{\iota}_*$, \autoref{fig:landscape1} also shows the landscape around the same $\bar{\iota}_*=0.42$ optimum after $\bar{\iota}_*$ in the objective has been changed to $0.43$, corresponding to the initial state of the optimization of the next step in our $\bar{\iota}_*$ scan. For some of the insensitive modes (for example, $m=3, n=0$), the objective changes from being relatively flat to having an incline over the plotted range. This will push the optimizer to make relatively large jumps in these boundary modes, potentially to different local minima in these insensitive modes.

Thus, the continuation method does not strictly follow a unique minimum as the landscape is deformed by changing $\bar{\iota}_*$. Rather, it samples a class of minima with potentially large relative differences in the insensitive modes.
In the next section we use the resulting equilibria to evaluate how various figures of merit vary with iota.





\subsection{Varying rotational transform in a quasi-axisymmetric equilibrium\label{sec:QAiota}}
\begin{figure}
  \includegraphics[width=\textwidth]{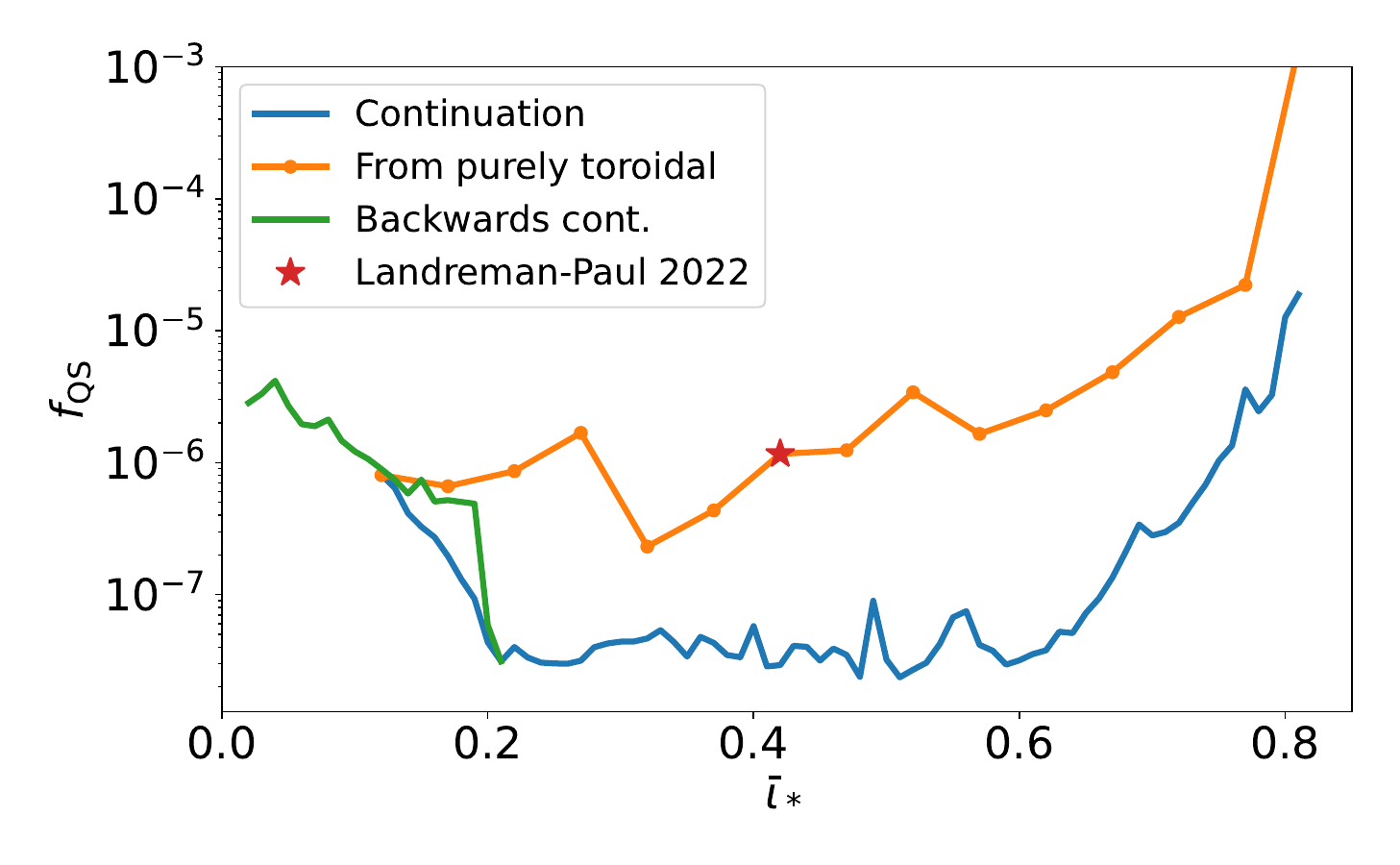}
  \caption{\label{fig:QAiota} Quasisymmetry error \eqref{eq:2term} from configurations with varying $\bar{\iota}$. The blue line corresponds to configurations found by the continuation method described in \autoref{sec:cp}. For comparison, we also include optimizations following the same procedure as \citet{landremanPaul2022}, but targeting different $\bar{\iota}$ (orange). In green is a continuation scan towards lower $\bar{\iota}$, showing that the scan is not reversible, which is expected given the ``non-uniqueness'' of the configurations as discussed in \autoref{sec:explore}. The ``Precise QA'' \citep{landremanPaul2022} corresponds to the $\bar{\iota}=0.42$ ``from purely toroidal'' optimization.}
\end{figure}

Using the objective function \eqref{eq:objective} with $A_*=6.0$, $N=1$ and $M=0$ in \eqref{eq:2term}, we optimize the boundary of a $n_{\text{fp}} = 2$ field-period vacuum configuration starting from a purely toroidal field and targeting $\bar{\iota}_* = 0.12$. We optimize by successively expanding the number of boundary Fourier modes in 4 steps as in \nouncite{landremanPaul2022}, only including Fourier modes up to $|n| \leq j, m \leq j$ in the $j$th step. This optimization is repeated for 6 different relative finite-difference step-sizes and the optimum with the lowest quasisymmetry error is picked. The resulting optimum is taken as our initial configuration in the procedure described in the next paragraph.

Incrementing $\bar{\iota}_*$ by $0.01$, we redo the optimization starting from the previous optimum. As this is a small adjustment to the objective function, the previous optimum is expected to be a good initial guess. Thus, we immediately optimize with all Fourier boundary modes. This not only saves time, but is also expected to make the resulting series of optima similar to each other. For convenience, all optimizations are performed with the initially found optimal finite-difference step size.

The quasisymmetry errors of the resulting configurations at the different $\bar{\iota}_*$ are plotted in \autoref{fig:QAiota}, alongside the results of rerunning the optimizations from a purely toroidal field at different $\bar{\iota}_*$. We see that the continuation method results in lower quasisymmetry error throughout the entire $\bar{\iota}_*$ range, achieving quasisymmetry errors about ten times lower than the optimization starting from a purely toroidal field. For $\bar{\iota}_* \in [0.2, 0.65]$, the quasisymemtry error displays a broad valley, where it varies between $2.4 \times 10^{-8}$ and $9.0 \times 10^{-8}$.

For $0.12 < \bar{\iota}_* < 0.2$, the continuation method gives decreasing quasisymmetry error for increasing $\bar{\iota}_*$. To test whether this is due to lingering effects of the initial configuration at $\bar{\iota}_* = 0.12$, we run the continuation method in reverse starting from the $\bar{\iota}_*=0.21$ configuration. The result is shown in green in \autoref{fig:QAiota}.
Due to the ``non-uniqueness'' property discussed in \autoref{sec:explore}, we do not expect the process to retrace itself exactly when done in reverse. Indeed, we find that the resulting configurations have somewhat different quasisymmetry error after only two backwards iterations, compared to the previously obtained configurations with the same $\bar{\iota}_*$. These new configurations are not notably better than the configurations obtained by starting from a purely toroidal field. This indicates that higher quasisymmetry error for $\bar{\iota}_* < 0.20$ is not merely an effect of the initial conditions in the optimization.

\begin{figure}
  \includegraphics[width=\textwidth]{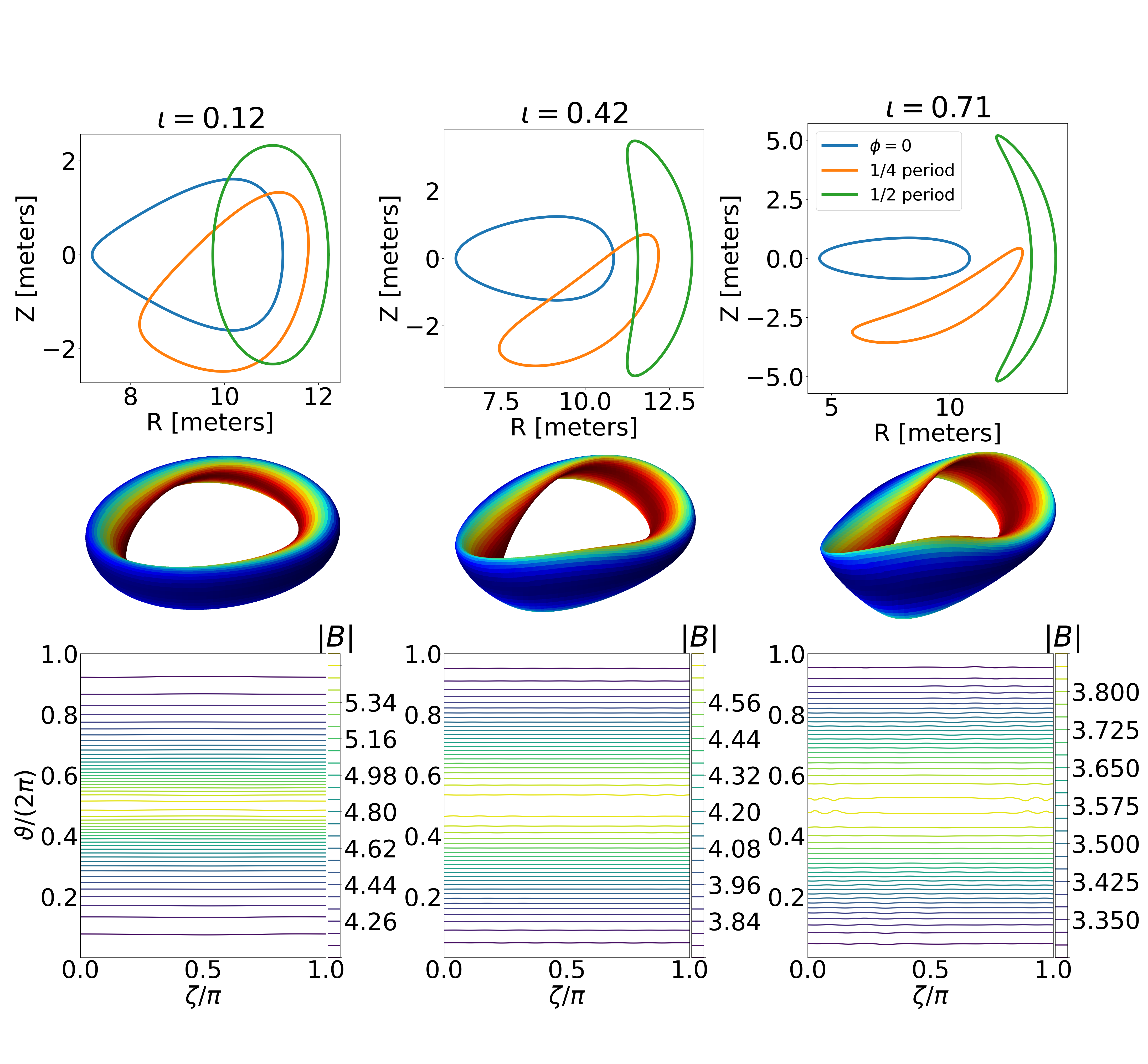}
  \caption{\label{fig:QAgeom} Row 1 \& 2: Plasma boundary shapes for select points in the scan ($\iota=0.12$, $\iota=0.42$ and $\iota=0.71$). As $\iota$ increases, the boundary becomes more elongated. Row 3: $|\vec{B}|$ on the boundary in Boozer coordinates.}
\end{figure}

\autoref{fig:QAgeom} shows examples of boundary shapes obtained in the continuation scan. We see that as $\bar{\iota}_*$ increases, the boundary shapes become more elongated. This appears to limit the range of achievable $\bar{\iota}_*$ in the scan, as VMEC eventually fails to solve for the resulting plasma equilibria around $\bar{\iota}_* = 0.81$. The quasisymmetry starts to degrade already at $\bar{\iota}_* \approx 0.65$, perhaps because of the optimum shape becoming thin enough that it cannot be accurately represented using only boundary Fourier modes up to $|n| \leq 4, m \leq 4$.


One advantage of higher $\bar{\iota}_*$ is that the orbit width $\Delta \psi$ of charged particles roughly scales as $1/\iota$ for QA configurations. Specifically, in the small orbit-width limit, the orbit width is proportional to \citep{paul2022}
\begin{equation}
   \Delta \psi  \propto \left|\frac{MG - NI}{M\iota - N}\right|.
\end{equation}
We thus expect fast particles to be better confined in QA configurations with higher $\bar{\iota}_*$, all else being equal. Note that orbit width in actual meters would have an even stronger dependence on $\iota$, since there is significant flux-surface compression for higher $\iota$.

\begin{figure}
  \includegraphics[width=\textwidth]{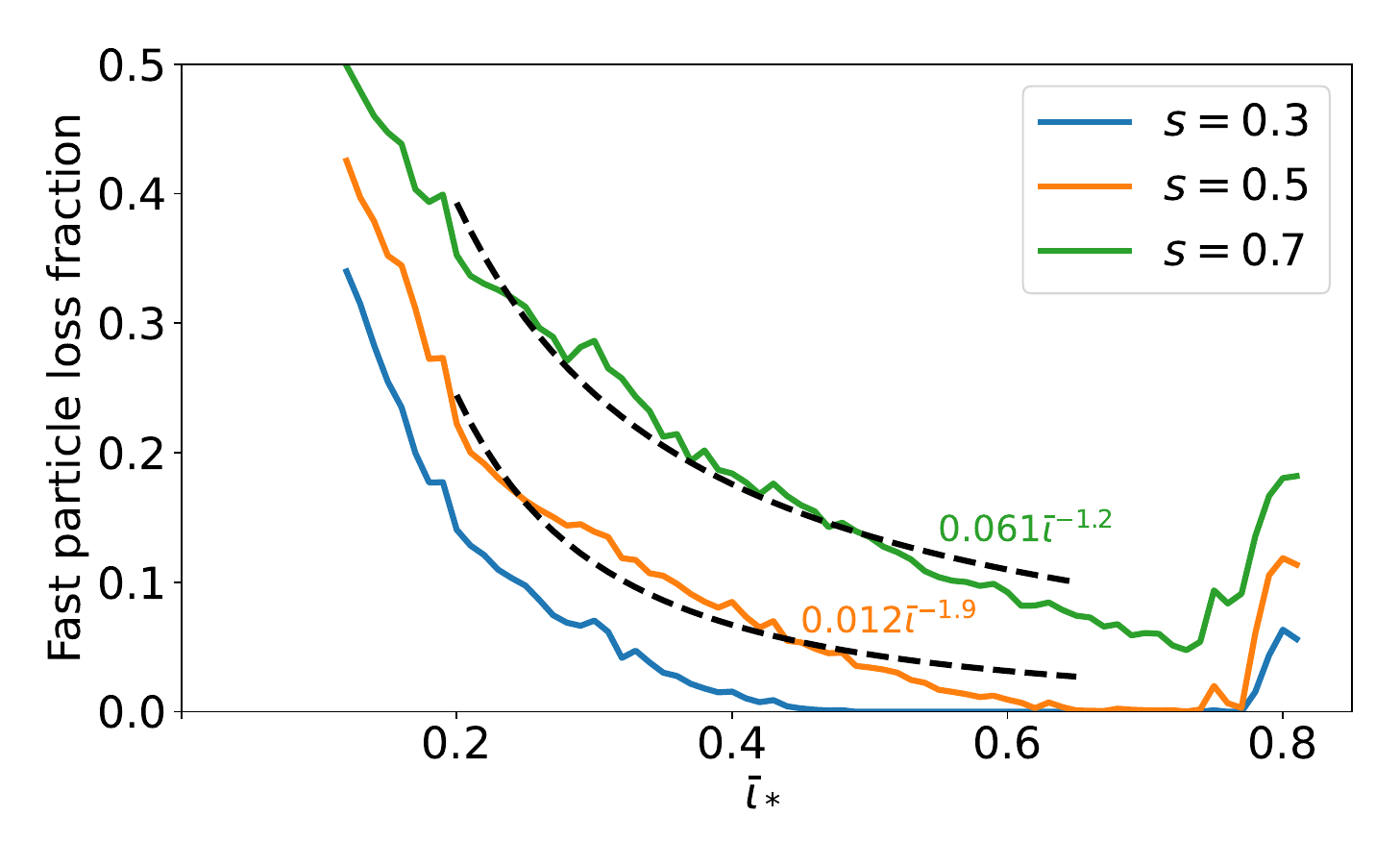}
  \caption{\label{fig:simpleQA}Fast-particle loss-fraction for the different $\bar{\iota}$ configurations as calculated by \textsc{Simple} by tracing $5000$ alpha-particles released at the surfaces with normalized toroidal flux $s=0.3$, $s=0.5$, $s=0.7$. The dashed lines show least-squares fits of $c\,\bar{\iota}_*^{\alpha}$ to the data, with fit parameters $c$ and $\alpha$ shown in the figure.}
\end{figure}

This prediction is borne out by particle tracing simulations using the collisionless guiding-center orbit solver \textsc{Simple} \citep{simple1,simple2}.
\autoref{fig:simpleQA} shows the loss fractions calculated by tracing $5000$ alpha particles with $3.52\,\mathrm{MeV}$ energy for $0.2$ seconds. The particles were launched at three different radii, and considered lost when crossing the last closed flux surface at normalized toroidal flux $s=1.0$. The configurations have all been scaled up to a reactor-relevant minor radius of $1.704\,\mathrm{m}$  and volume average magnetic field $5.865\,\mathrm{T}$, to match the ARIES-CS configuration \citep{ariescs2008}.
We see that the confinement improves with $\bar{\iota}$ until about $\bar{\iota} = 0.73$, where the increase in quasisymmetry error out-weights the benefits of increasing $\iota$.

To see to which extent the $1/\iota$ orbit-width scaling is reflected in the fast-particle losses, we fit the losses to $c\bar{\iota}_*^{\alpha}$ for $\bar{\iota}_* \in [0.2,0.65]$, where the quasisymmetry error is relatively constant. The fits and values of the coefficients are shown alongside the data in \autoref{fig:simpleQA}, for the $s=0.5$ and $s=0.7$ results. The fits systematically overpredict the particle losses towards the higher end of the fitted range.
Increasing the range of $\bar{\iota}_*$ in the fit to $[0.12,0.73]$ changes of the values of the coefficients to ($c=0.07$, $\alpha=-0.99$) and ($c=0.017$, $\alpha=-1.6$) for the $s=0.7$ and $s=0.5$ fits, respectively, and does not significantly affect the agreement between the fitted curve and the data. Thus, the orbit width scaling with $\iota$ only offers a qualitative estimate of how the fast-particle losses scale, and other mechanisms must be accounted for to get the full picture.

Not included in the \textsc{Simple} calculations is the effect of the gyro-orbit about the guiding-centers, which effectively widens the orbits and thus should increase the losses.
As an estimate of the size of this effect, we calculate the ratio of the gyroradius to the orbit width, using the expression for the orbit width in the small orbit-width limit given by \citet{paul2022}
\begin{equation}
\frac{\text{gyroradius}}{\text{orbit width}} =     \frac{|N - \iota M|}{|MG - NI|}|\nabla \psi| \frac{\sqrt{B\lambda}}{2\sqrt{1-B\lambda}}.
\end{equation}
For our QA configurations ($I\approx 0\,\mathrm{Tm}$, $G\approx 70\,\mathrm{Tm}$, |$\nabla \psi| \approx 20\,\mathrm{Tm}$) this ratio ends up being roughly $\iota / 7$ for $B\lambda = 0.5$, and is thus much less than one. The finite gyro-orbit correction should thus be small for our range of $\iota$.


\begin{figure}
  \includegraphics[width=\textwidth]{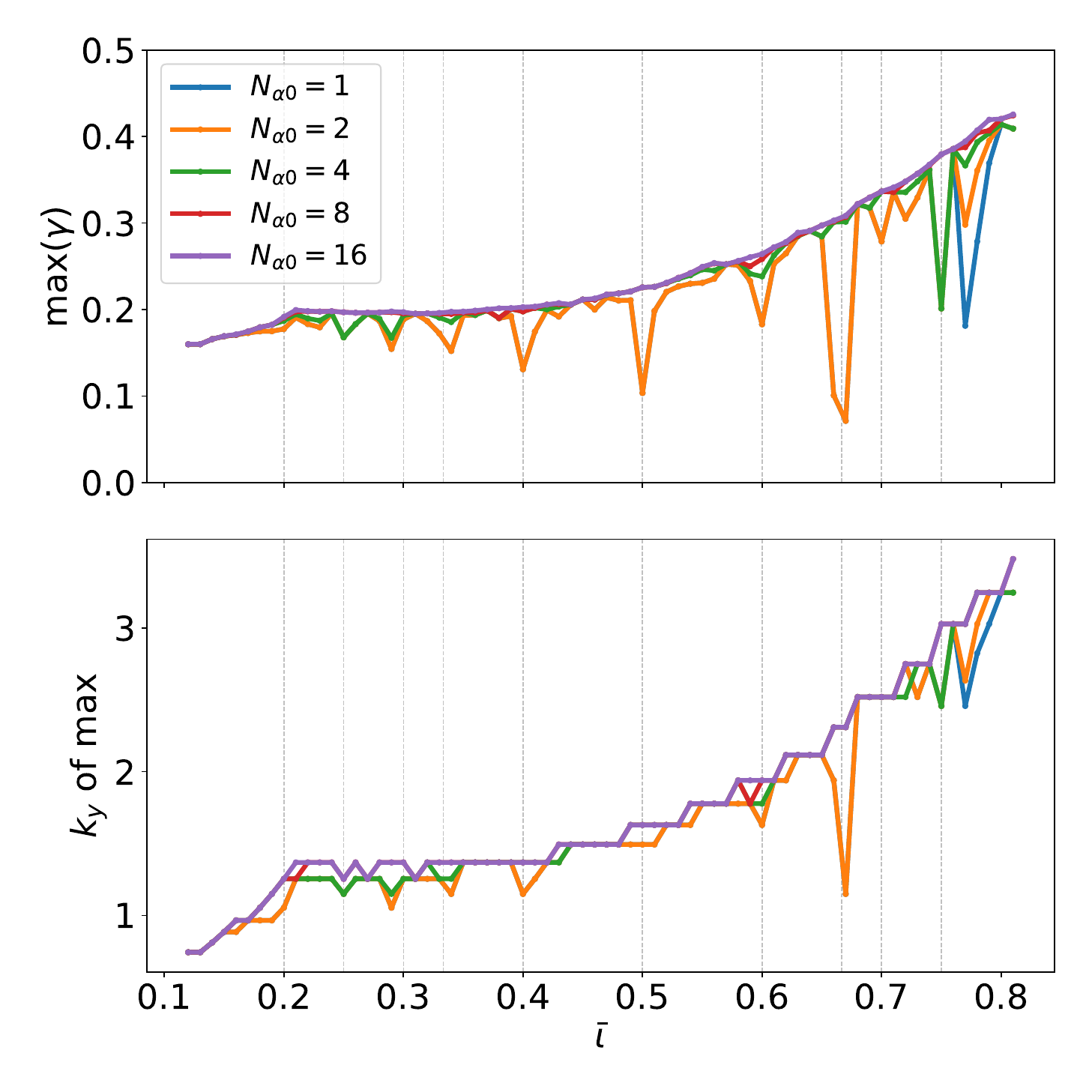}
  \caption{\label{fig:gammaQA} (a) Maximum growth rate at $s=0.25$, for a different number $N_\alpha$ of flux tubes in each $\bar{\iota}$ configuration. Data are from electrostatic linear \textsc{Stella} simulations with $a/L_{T} = 3$, $a/L_n = 1$ and adiabatic electrons. (b) The $k_y$ mode-number of the fastest growing mode. The underlying $\gamma(k_y)$ curves are shown in \autoref{fig:gamma2QA}. Vertical dotted lines indicate some low-order rationals.}
\end{figure}

\begin{figure}
  \includegraphics[width=\textwidth]{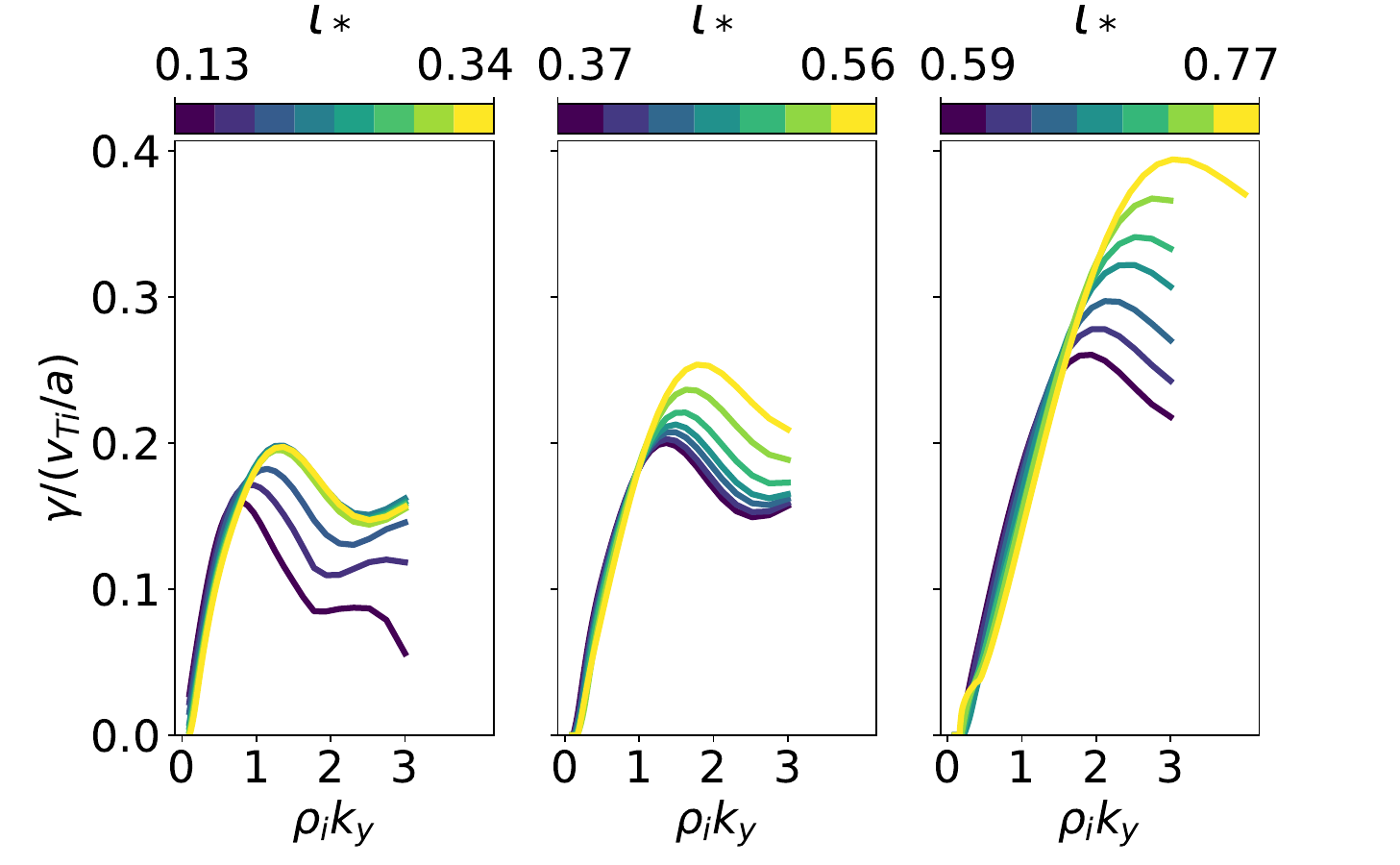}
  \caption{\label{fig:gamma2QA} Growth rates $\gamma(k_y)$ for the flux tube with the highest $\max{(\gamma)}$ for a subset of the different $\bar{\iota}_*$ configurations. The curves for different configurations are grouped into different subfigures based on the range of their $\bar{\iota}_*$ values: (a) 0.13-0.34, (b) 0.37-0.56, (c) 0.59-0.77. Colors indicate the specific $\bar{\iota}_*$ corresponding to each curve.}
\end{figure}

For configurations with this low quasisymmetry error, collisional transport is almost surely dwarfed by turbulent transport. For example, ion-temperature gradient (ITG) mode turbulence appears to dominate the heat transport in Wendelstein 7-X \citep{carralero2021}.

In \autoref{fig:gammaQA}, we show the maximum growth rates $\max{(\gamma)}$ for the $s=0.25$ surface in each configuration, as calculated from linear electrostatic gyrokinetic flux-tube simulations using the gyrokinetic code \stella{} \citep{stella1} using adiabatic electrons.
The simulations use gradients to promote ITG turbulence: $a/L_{Ti} = 3$ and $a/L_{n} = 1$.
In the figure, $N_\alpha$ indicates the number of flux tubes in which $\max{(\gamma)}$ was calculated, using equally spaced values of the field line label $\alpha_0$ on the interval $[0, \pi]$, with a point always placed on $\alpha_0 = 0$. In \stella{}, $\alpha_0$ (together with $s$) specifies the location of a flux tube, with $\alpha_0$ being the value of $\alpha = \theta - \iota \phi$ at the middle of the flux tube. The $\alpha_0$ interval $[0, \pi]$ covers the entire half-period of a stellarator symmetric configuration. The simulations use $k_x=0$ and the parallel domain extends for 70 toroidal turns.

The $\gamma(k_y)$ curves corresponding to the flux tube with the maximum growth rates are shown in \autoref{fig:gamma2QA}, for some of the $\bar{\iota}_*$ configurations. Despite some of the curves being calculated on flux tubes with different $\alpha_0$, the result is an essentially smooth change in the growth rate curves with $\bar{\iota}_*$. For $\bar{\iota}_* \ge 0.77$, the growth rate peaks above $\rho k_y = 3.0$, so these simulation were modified to include higher $k_y$ values. 

\begin{figure}
  \includegraphics[width=\textwidth]{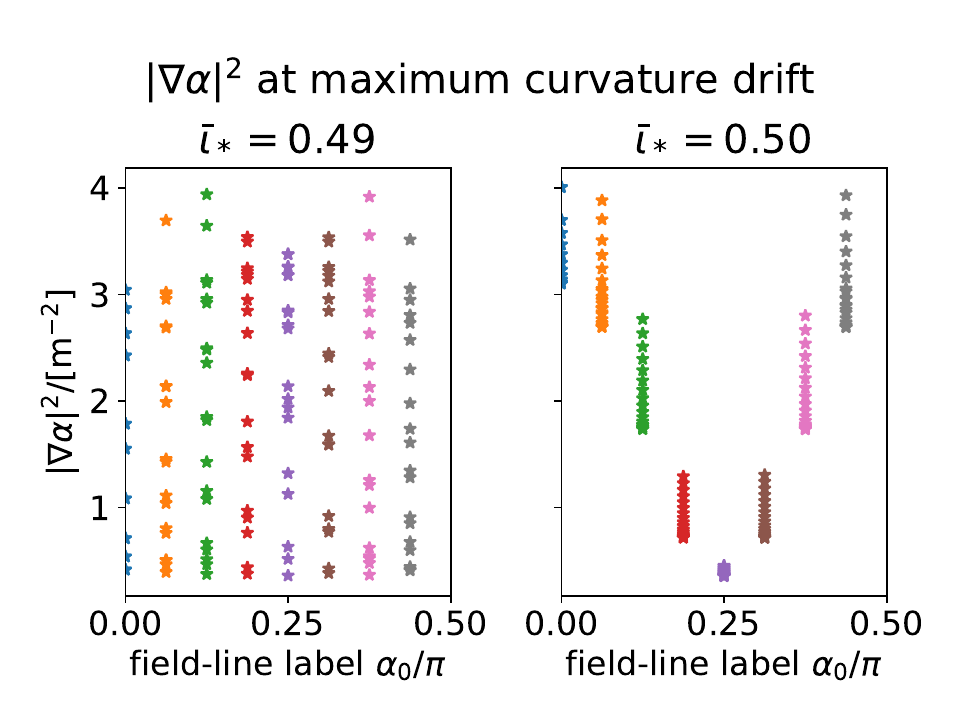}
  \caption{\label{fig:gds2scatter} Values of $|\nabla \alpha|^2$ at points of maximum curvature drift along different flux tubes (given by $\alpha_0$ on the x-axis) for an irrational ($\bar{\iota}_* = 0.49$ and rational ($\bar{\iota}_* = 0.50$) flux surface.}
\end{figure}

\autoref{fig:gammaQA} and \autoref{fig:gamma2QA} both show a clear trend towards higher $\max{(\gamma)}$ and higher wave-numbers $k_y$ with increasing $\bar{\iota}$, suggesting smaller scale turbulence. At certain values of $\bar{\iota}$, the $N_{\alpha 0} =1$ and $N_{\alpha 0} =2$ curves in \autoref{fig:gammaQA} display sharp dips towards lower $\max{(\gamma)}$.

This can be understood in terms of the flux tubes on rational surfaces not sampling the full flux-surface geometry. As a result, modes on different flux tubes will (for example) experience different levels of finite-Larmor-radius (FLR) damping in regions of bad curvature. This is illustrated in \autoref{fig:gds2scatter}, which shows the value of $|\nabla \alpha|^2$ at maxima in the curvature drift along the flux tube, for different flux tubes in the  $\bar{\iota}_* = 0.49$ and  $\bar{\iota}_* = 0.50$ geometries. For the non-rational $\iota$, all flux tubes sample regions of bad curvature with all values of $|\nabla \alpha|^2$, while for the rational $\iota$, the different flux tubes display different values of $|\nabla \alpha|^2$ at regions of bad curvature. Flux tubes with lower $|\nabla \alpha|^2$ in bad curvature regions have higher growth rates as they experience less FLR damping.


\begin{figure}
  \includegraphics[width=\textwidth]{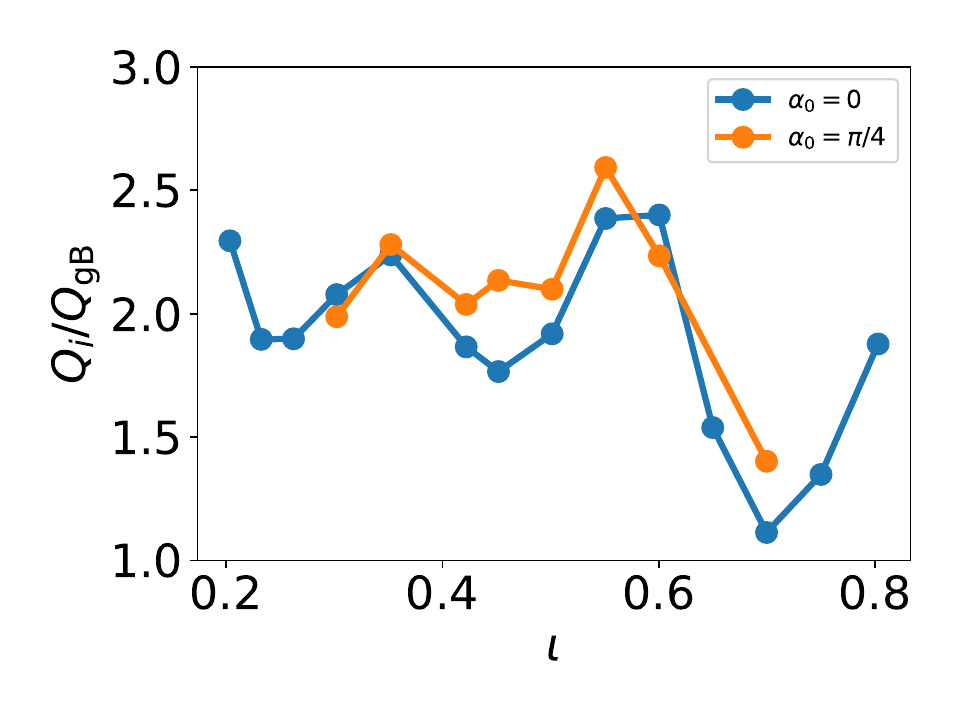}
  \caption{\label{fig:NLQi} Nonlinear ion heat fluxes $Q_i$ for a few different $\bar{\iota}_*$ configurations starting from $\bar{\iota}_*=0.30$ to $\bar{\iota}_*=0.75$. The fluxes are calculated using the gyrokinetic code \textsc{Gx} on a single flux tube, with calculations done for the $\alpha_0 = 0$ and $\alpha_0=\pi/4$ flux tubes.
  }
\end{figure}

\autoref{fig:NLQi} shows the corresponding nonlinear ion heat flux as calculated using the gyrokinetic code \GX{}. \GX{} is a new gyrokinetic code that runs on GPUs  \citep{mandell2022gx, mandell2018}. \GX{} has been benchmarked against \stella{} and other established gyrokinetic codes \citep{mandell2022gx}, and should be faster than \stella{} when it comes to running electrostatic nonlinear gyrokinetic simulations with adiabatic electrons. 
In contrast to the linear results, there is more variation between flux tubes with different $\alpha_0$, but no great difference between rational and irrational values of $\iota$. Both these features are consistent with the nonlinear heat flux being insensitive to structures that are very elongated along the field-line, so that even on irrational flux tubes, the turbulence does not effectively sample all regions of bad curvature on the flux surface.
The trend with respect to $\iota$ is non-monotonic with several local maxima and minima.


\begin{figure}
\centering
\includegraphics[width=\textwidth]{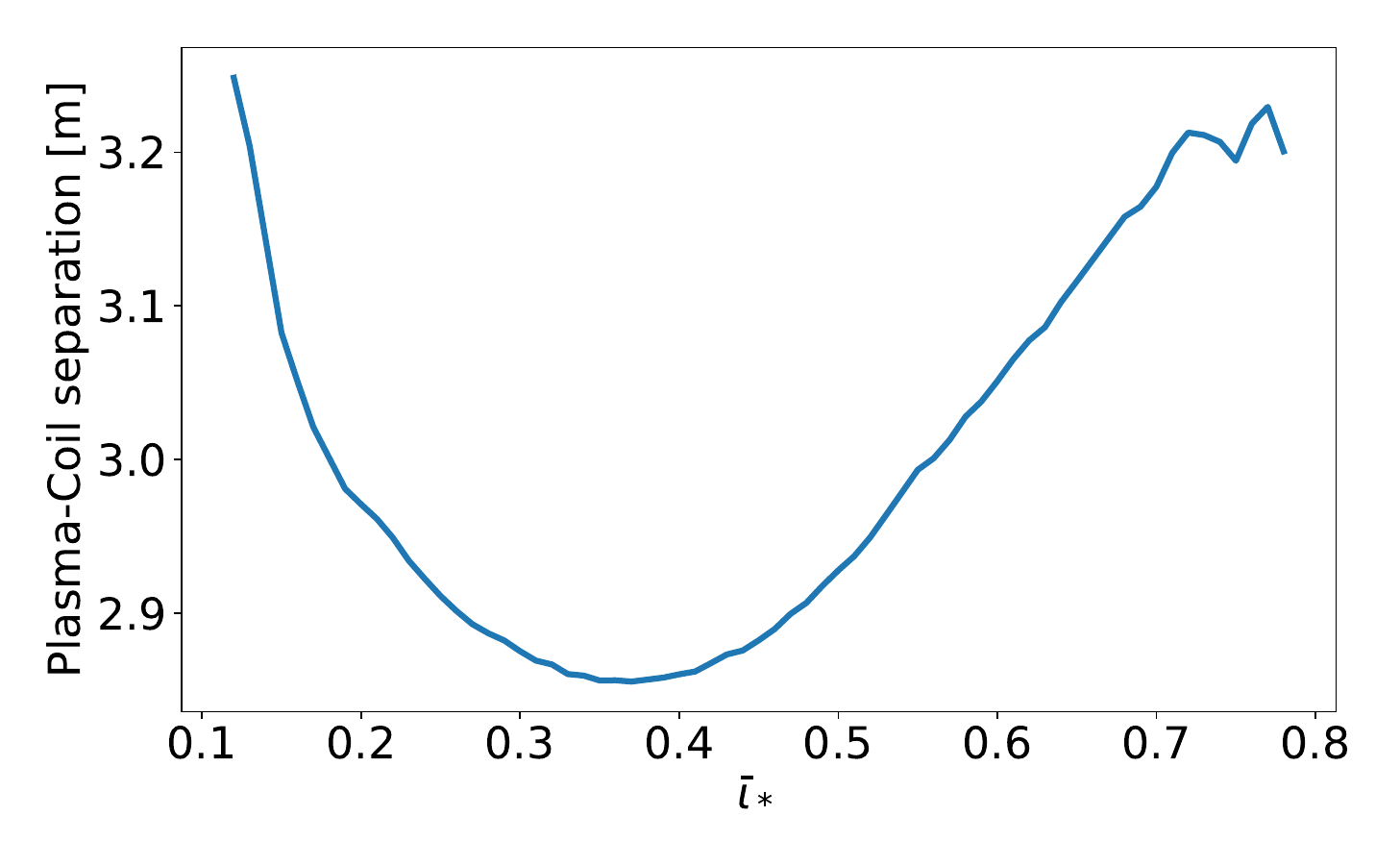}
\caption{\label{fig:plasma_coil_sep}Coil-plasma separation for the different $\bar{\iota}_*$ configurations, calculated using \textsc{Regcoil}. The configurations have been scaled up to $a=1.704\,\mathrm{m}$  and volume average magnetic field $5.865\,\mathrm{T}$, to match ARIES-CS. 
The sheet current density is computed to meet a target accuracy $B_{\text{n,RMS}} = 0.01\,\mathrm{T}$ and a maximum current density of $17.16\,\mathrm{MA/m}$, which corresponds to the minimum coil-coil distance of ARIES-CS \citep{kappel2023prep}.}
\end{figure}

We also investigated how far away coils can be placed from our plasma, using the method in \citet{kappel2023prep}. Given a maximum surface-current density $K_{\text{max}}$ (corresponding to a minimum coil-coil distance) and a required accuracy in reproducing the last-closed flux-surface of a target configuration, the distance between the plasma and the coil winding surface in the coil-shape code \textsc{Regcoil} \citep{regcoil1} is increased until the maximum current density on the winding surface reaches $K_{\text{max}}$.
The accuracy with which the coils produce the target last-closed flux-surface is measured by
\begin{equation}
B_{\text{n,RMS}} = \sqrt{\int_{\text{LCFS}} \d S \left(\vec{n} \cdot \vec{B}\right)^2/\int_{\text{LCFS}} dS},
\end{equation}
where the surface integrals are over the target surface and $\vec{n}$ is the unit normal of this surface; $\vec{B}$ is here the magnetic field produced by the coils.

The results of \textsc{Regcoil} calculations with $K_{\text{max}}=17.16\,\mathrm{MA/m}$ and $B_{\text{n,RMS}} = 0.01\,\mathrm{T}$ for the different $\bar{\iota}_*$ target configurations are shown in \autoref{fig:plasma_coil_sep}. All target configurations have been rescaled to the ARIES-CS minor radius $a$ and volume averaged $B$, and the value of $K_{\text{max}}$ was chosen to match the minimum coil-coil distance of the ARIES-CS coil-set.
From the figure, we see that the coil-plasma separation is between $2.85\,\mathrm{m}$ and $3.25\,\mathrm{m}$  with a minimum for $\bar{\iota}_*$ around $0.35$. These values are well above the estimated distance of $1.5\,\mathrm{m}$ to $2.0\,\mathrm{m}$ required to fit the scrape-off layer, first wall, breeding blanket and neutron shielding \citep{ariescs2008}.

\subsection{Varying shear in quasi-axisymmetric equilibrium}
All configurations in the previous section have low values of global magnetic shear. This might make them susceptible to curvature driven instabilities, compared to configurations with negative shear \citep{antonsen1996, nadeem2001}. 

High shear helps limiting the parallel extent of modes elongated along the magnetic field lines due to radially nearby field lines separating as the parallel coordinate is traversed, which forces such modes to have lower radial correlation length or a limited parallel extent. Gyrokinetic modes with smaller perpendicular extent generally have a smaller impact on a configuration, as they lead to less transport, which can be qualitatively understood by viewing the size of the mode perpendicular to the field line as a step-length in a diffusive process \citep{merzPRL2008}.

We obtain configurations with stronger shear by additional optimizations where we add a term targeting the mean shear. Defining the mean shear as
\begin{equation}
  \bar{\hat{s}} = -c_1/\bar{\iota}, \label{eq:ourshear}
\end{equation}
where $c_1$ is the result of a linear fit $c_0+c_1 s$ to the $\iota(s)$ profile, we add the following optimization target to our objective \eqref{eq:objective}:
\begin{equation}
   (\bar{\hat{s}} - \bar{\hat{s}}_*)^2 \label{eq:shearobjective}.
 \end{equation}
 The negative sign in \autoref{eq:ourshear} is included to match the sign convention of the conventional definition of shear $-\frac{s}{\iota}\frac{\d \iota}{\d s}$.

\begin{figure}
  \includegraphics[width=\textwidth]{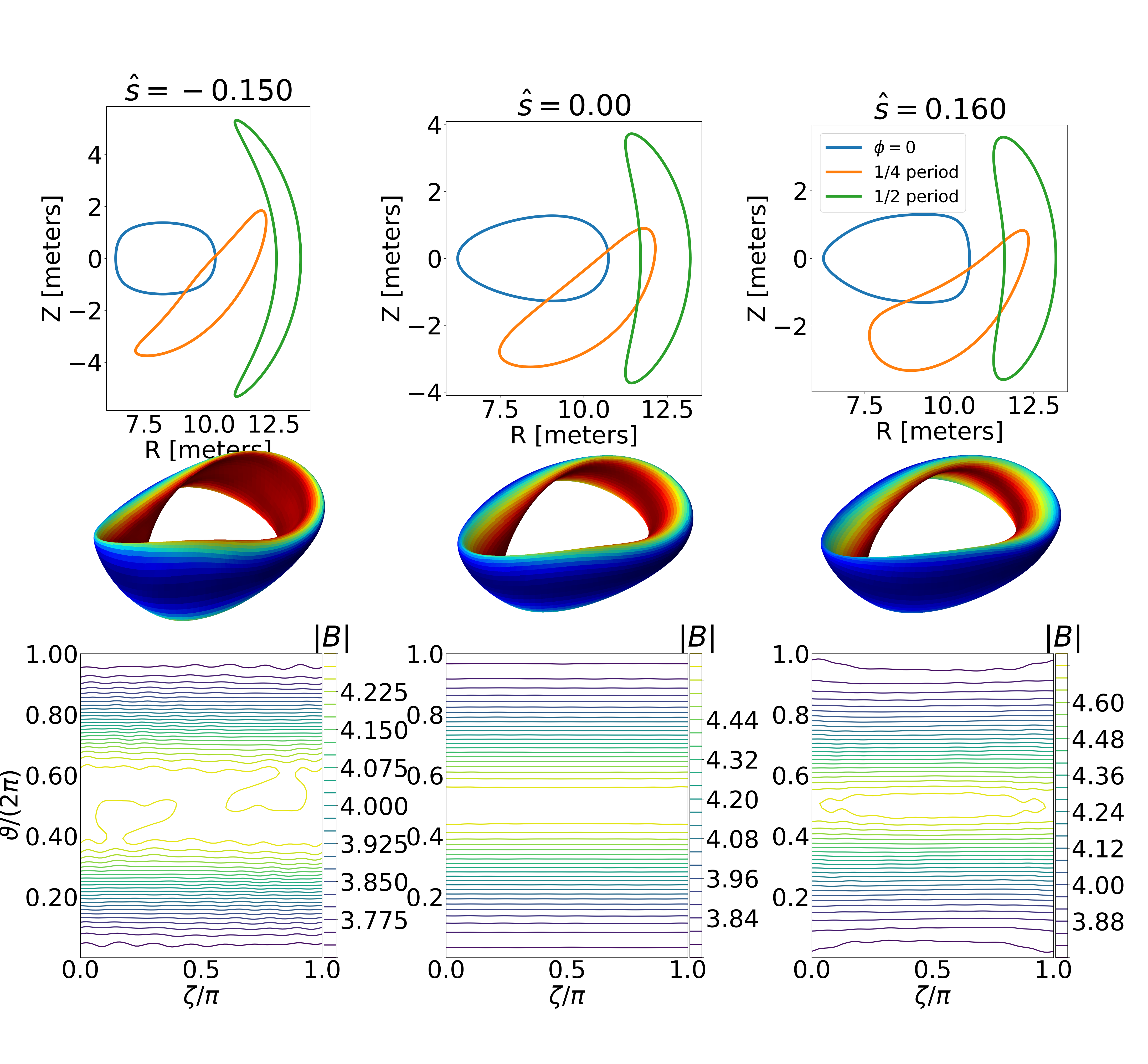}
  \caption{\label{fig:QAsheargeom} Plasma boundary shapes for select points in the scan ($\bar{\hat{s}}=-0.16$, $\bar{\hat{s}}=0.0$ and $\bar{\hat{s}}=0.15$). Imposing positive shear mainly affects the triangularity of the boundary, while a negative shear makes the boundary more elongated and narrow.}
\end{figure}

Starting from the $\bar{\iota}_* = 0.42$ configuration found in \autoref{sec:QAiota}, which has a mean shear  $\bar{\hat{s}}  = 0.0195 \approx 0.02$, we do two continuation scans in $\bar{\hat{s}}_*$ towards more negative and more positive shear, respectively, still targeting $\bar{\iota}_* = 0.42$. The configurations with lowest and highest shear are shown \autoref{fig:QAsheargeom}, alongside the $\bar{\hat{s}} = 0$ configuration.

\begin{figure}
  \includegraphics[width=\textwidth]{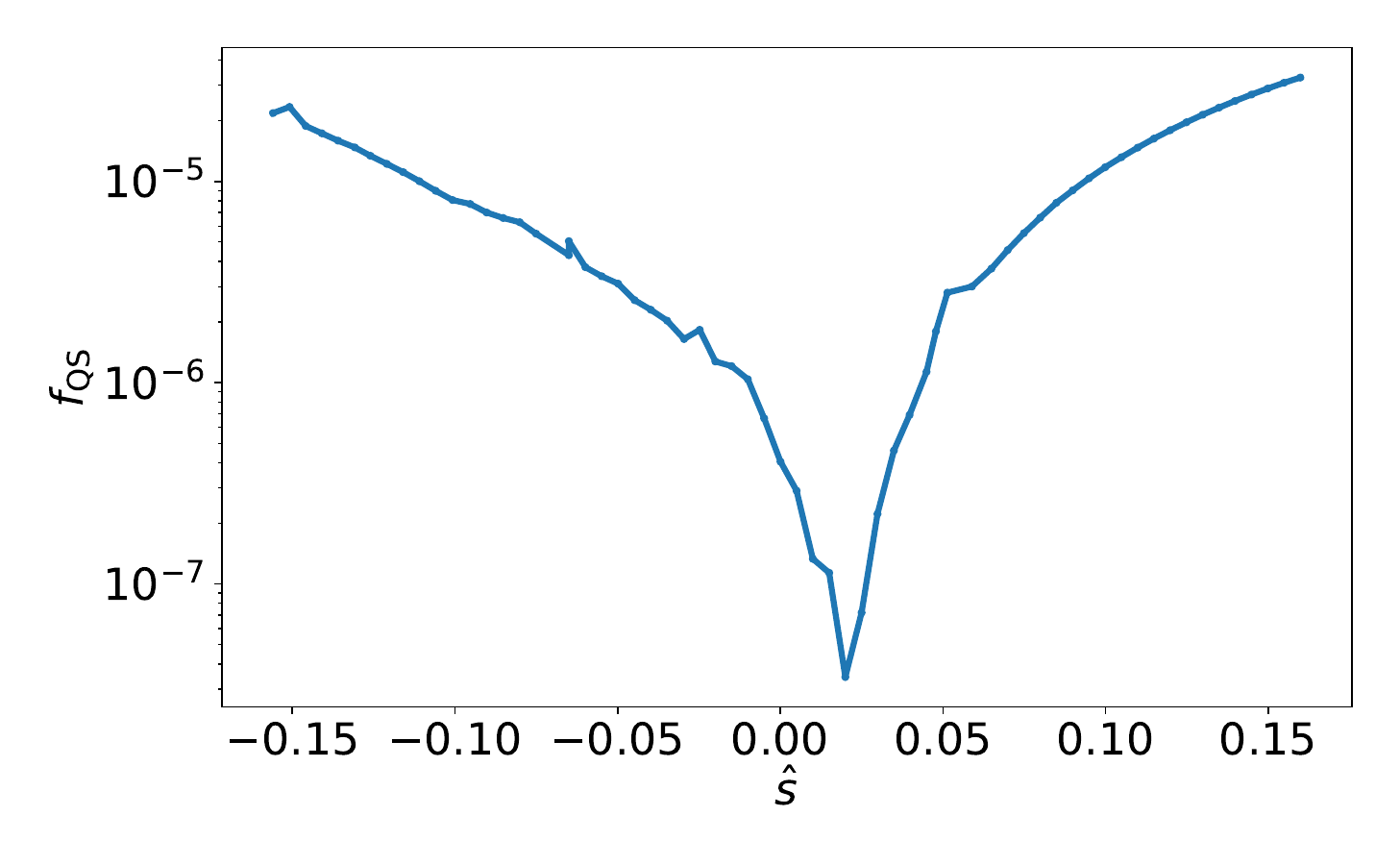}
  \caption{\label{fig:QAshear} Quasisymmetry error \eqref{eq:2term} from a continuation scan varying shear. The scan is performed around the $\iota=0.42$ configuration.}
\end{figure}

The resulting quasisymmetry error is shown in \autoref{fig:QAshear}. We see that imposing an additional shear objective degrades the quasisymmetry. This is expected when adding competing objectives to an optimization problem.

\begin{figure}
  \includegraphics[width=\textwidth]{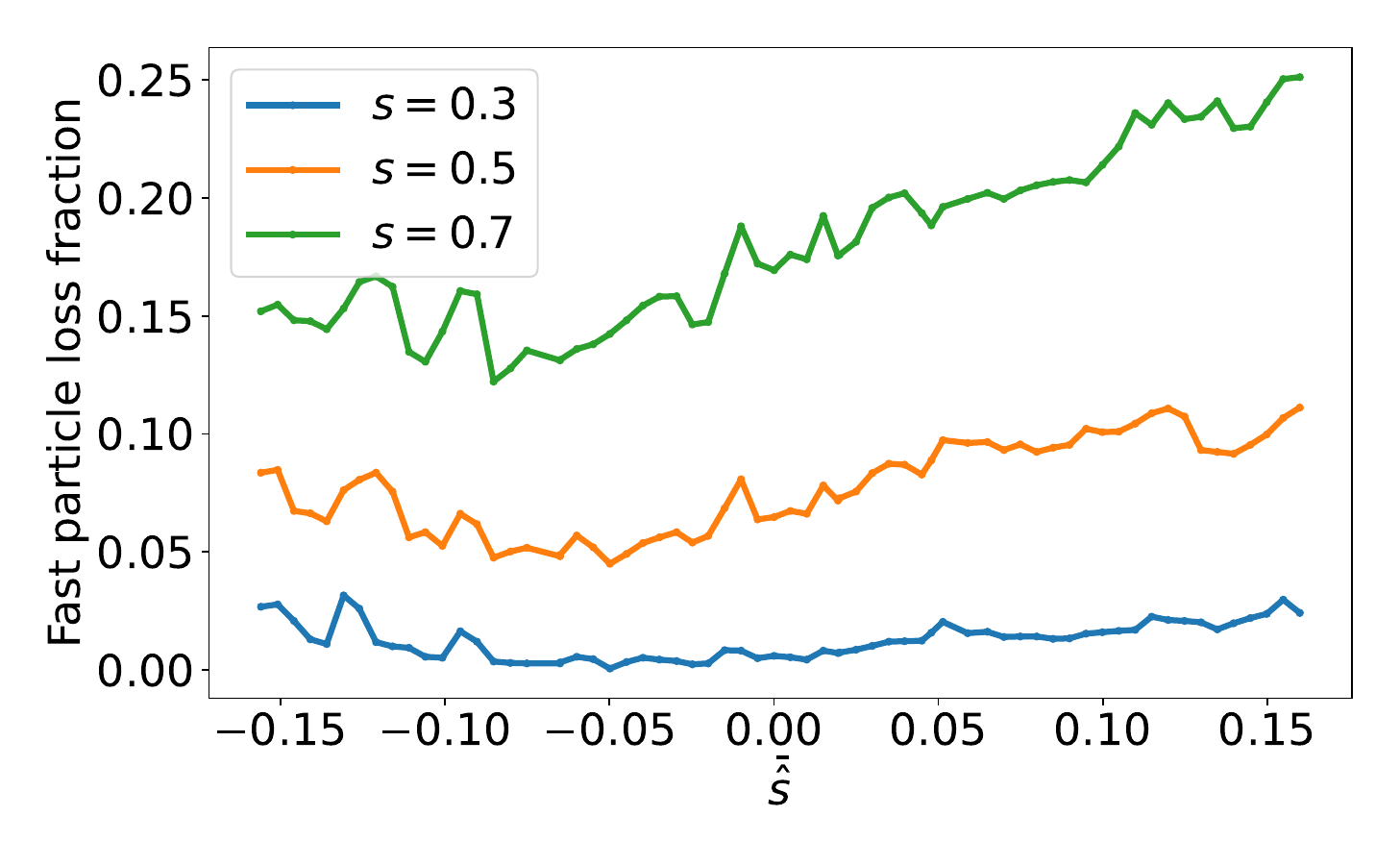}
    \caption{\label{fig:QA2shear} Fast-particle loss-fraction for the different $\bar{\hat{s}}$ configurations as calculated by \textsc{Simple} using the same simulation setup as in \autoref{fig:simpleQA}.}
  \end{figure}

  In \autoref{fig:QA2shear}, we show the fast-particle loss-fraction calculated using the same setup as the calculations in \autoref{fig:simpleQA}. Despite the order-of-magnitude difference in quasisymmetry error at the different values of $\hat{s}$, the losses only varies by at most $\pm 25\%$ (for $s=0.7$) and do not seem to be strongly correlated with the quasisymmetry quality. Thus, the entire $\hat{s}$ range likely has good enough quasisymmetry that other factors play a more a important role in determining the fast-particle confinement.

  \begin{figure}
  \includegraphics[width=\textwidth]{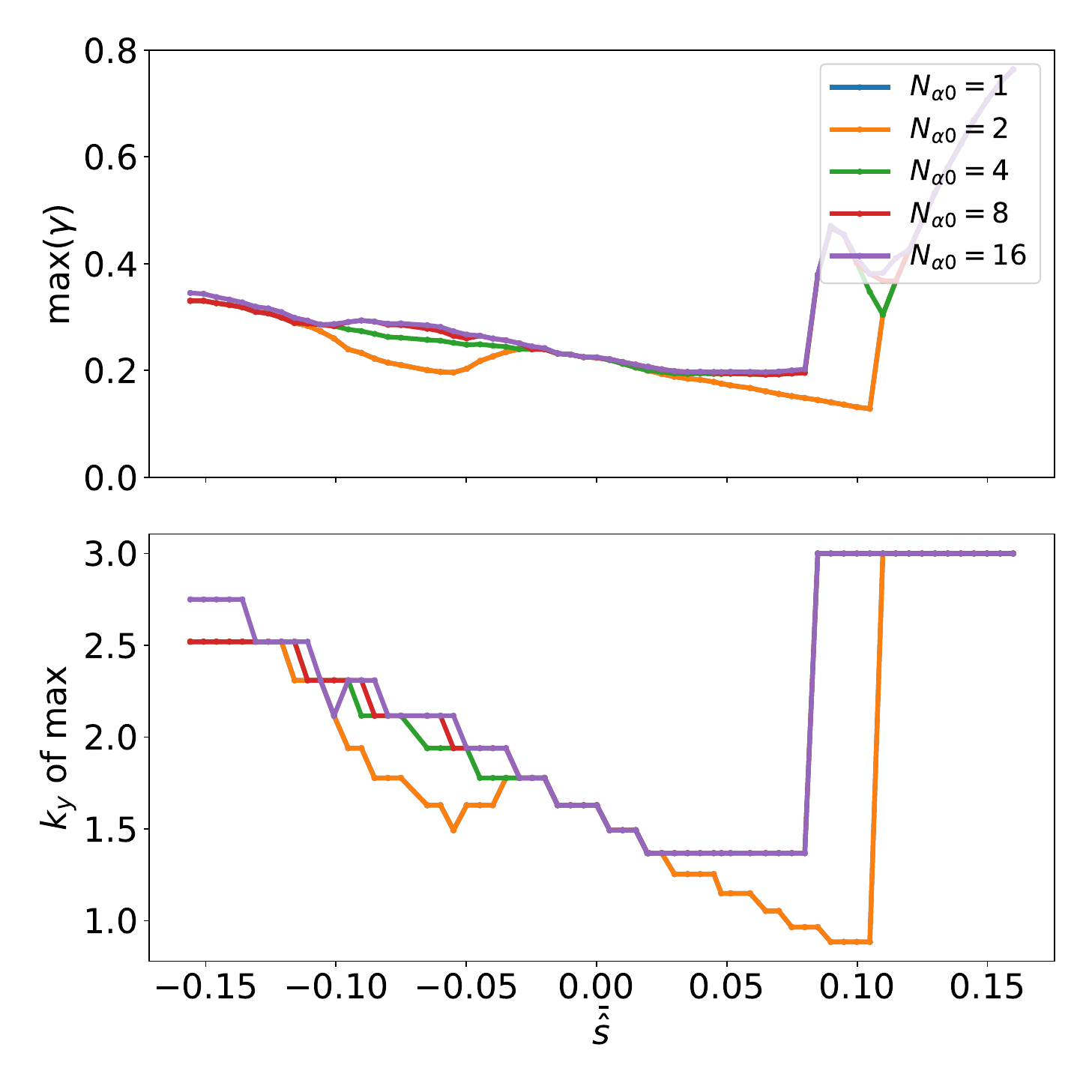}
    \caption{\label{fig:QA3shear} Maximum growth rates for the different $\bar{\hat{s}}$ configurations,  using the same simulation setup as in \autoref{fig:gammaQA}.}
  \end{figure}

    \begin{figure}
  \includegraphics[width=\textwidth]{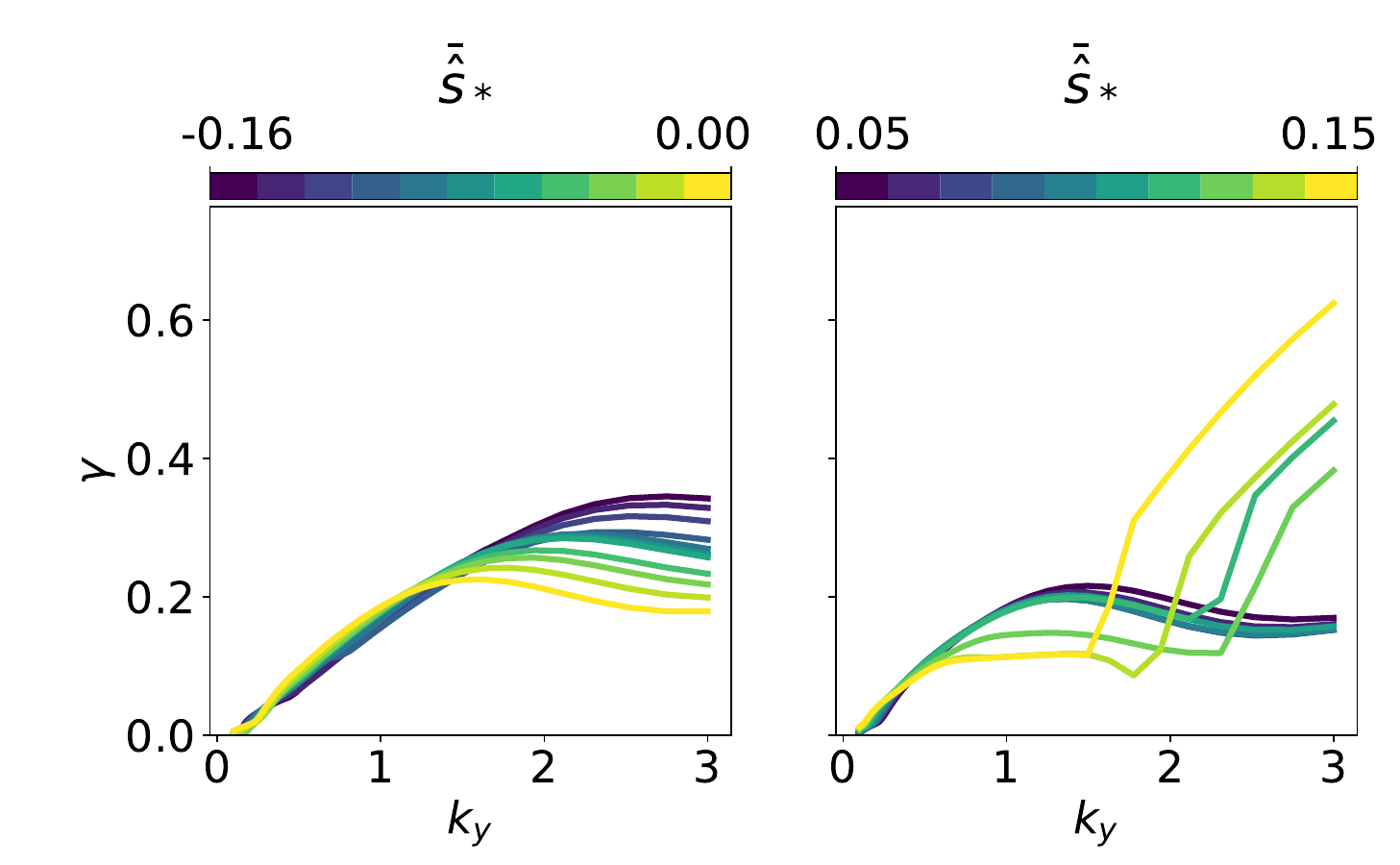}
    \caption{\label{fig:QA4shear}  Growth rates $\gamma(k_y)$ for a subset of the different $\bar{\hat{s}}_*$ configurations.}
  \end{figure}

\begin{figure}
    \includegraphics[width=\textwidth]{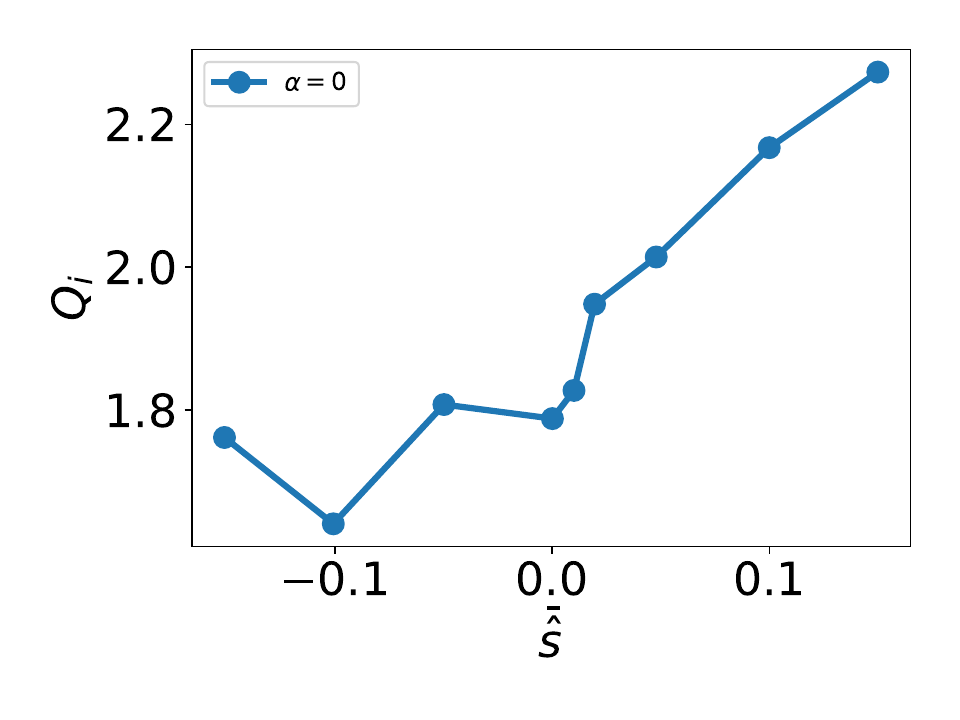}
    \caption{\label{fig:QA5shear}  Nonlinear heat fluxes for the different $\bar{\hat{s}}$ configurations.}
  \end{figure}

  Finally, we investigate the effect of shear on turbulent transport.
  The maximum growth rates $\max{(\gamma)}$ and and the dependence of the growth rate on $k_y$ for the shear scan are shown in \autoref{fig:QA3shear} and \autoref{fig:QA4shear}, respectively. We see a clear trend towards increasing maximum growth rates for shear above $\bar{\hat{s}} > 0.085$. This increase seems to be due to some new mode appearing at high $k_y$ values, and the maximum growth rate of this mode is not properly resolved in these simulations. The mode remains even when setting the density gradient to zero, and must thus be driven by the ion temperature gradient, although it appears at much larger values of $k_y$ than typical ITG turbulence.
The effect on the nonlinear ion heat flux is more modest, but there is an increase of about $25\%$ for positive shear, as seen in \autoref{fig:QA5shear}.
  
\section{Conclusions\label{sec:discussion}}
We have used a continuation method to generate a sequence of quasiaxisymmetric (QA) equilibria with different rotational transform profiles.
The configurations are similar to each other, with each a local minimum of an optimization problem that is initialized at the previously found minimum.
Through this procedure, we have found $n_{\text{fp}} = 2$ QA configurations with higher mean rotational transform $\bar{\iota}$ than what was previously found. We find QA configurations with record low quasisymmetry errors for $\bar{\iota} \in [0.2, 0.65]$. For $\bar{\iota}$ in and slightly above this range, increasing $\bar{\iota}$ increases the elongation of the plasma boundary, which is likely what ultimately limits the achievable values of $\bar{\iota}$. Fast-particle losses as calculated by a collisionless drift-kinetic particle-tracing code decrease with $\iota$ up to a point, but the effect is weaker than the $\iota^{-1}$ scaling predicted by purely accounting for how the orbit widths scales with $\iota$. Nevertheless, there is a potential conflict between having low elongation and low fast-particle losses.
The dependence of the turbulent ion heat flux is non-monotonic in $\iota$ and shows some mild sensitivity to which flux tube the simulations are performed at. The linearly calculated growth rates, on the other hand, are sensitive to the choice of flux tube on rational flux surfaces.

Higher shear can be imposed on the configurations by adding an extra term in the objective function. This increases the quasisymmetry error somewhat, but this hardly affects the fast particle losses. The turbulent heat flux tends to increase somewhat for positive shear, causing an increases by about $25\%$ for the range of shear considered here. The effect of shear in magnetohydrodynamic stability is likely even more important, but was not considered here.


\section*{Acknowledgements}

The authors are grateful for discussions with Bill Dorland, Ian Abel, Rogerio Jorge, Eduardo Rodriguez, Wrick Sengupta, Nikita Nikulsin, Richard Nies, Toby Adkins, and Jason Parisi, who have all provided valuable input and motivation for this work.

\section*{Funding}

This work was supported by the U.S. Department of Energy, Office of Science, Office of Fusion Energy Science, under award number DE-FG02-93ER54197.

\bibliographystyle{jpp}

\bibliography{plasma}

\end{document}